\documentclass[floats,aps,twocolumn]{revtex4}
\usepackage{calc}
\usepackage{graphicx}
\begin{document}
\def \beq{\begin{equation}}
\def \eeq{\end{equation}}
\begin{abstract}
We study meron quasiparticle excitations in the $\nu = 1$ quantum Hall bilayer.
Considering the well known single meron state, we introduce its effective
 form, valid in the longdistance limit. That enables us to propose two (and more) meron
states in the same limit. Further, establishing a plasma
analogy of the (111) ground state, we find the impurities that play the role
of  merons and derive meron charge distributions. 
Using the introduced meron constructions in  generalized (mixed) ground states
and corresponding plasmas for arbitrary distance between the layers,
we calculate the interaction between the construction implied impurities. We also find a 
correspondence between  the impurity interactions
and meron interactions. This suggests a possible explanation of the
deconfinement of the merons recently observed in the experiments.
\end{abstract}
\title{Meron excitations in the $\nu = 1$ quantum Hall bilayer and the plasma analogy}
\author{Milica V. Milovanovi\'{c} and Ivan Stani\'{c}}
\address{Institute of Physics, P.O.Box 68, 11080 Belgrade, Serbia and Montenegro}
\date{\today}
\maketitle
\vskip2pc]
\narrowtext
The $\nu = 1$ quantum Hall bilayer \cite{old}
 is the subject of intensive experimental and theoretical
investigations \cite{eimac}. In 1995, the pseudospin theory
of the bilayer \cite{moo} was advanced for this system. The theory 
introduced meron - a new type of quantum Hall quasiparticle. Nevertheless,
even today it is not known how a construction of a pair of merons looks  \cite{ye}.
Understanding of that would bring us closer to the understanding of
increased susceptibility to the presence of disorder of the neutral superfluid
in the pseudospin channel of the bilayer. Namely because of the persistent
dissipation in the counterflow measurments \cite{cal,pri}, there is
a wide-spread belief that even in the presence of a moderate amount of
disorder, merons -- vortices of the superfluid are liberated, dissociated
from one another \cite{she,hus,eimac}. On the other hand, since 
Laughlin's seminal paper \cite{lau}, the plasma analogy has proven to be
a very useful concept in analyzing quasiparticle state properties.

In this paper we develop a description of the meron excitations of the
pseudospin theory in the longdistance limit. We use the plasma techniques of
\cite{meth}, and argue that meron presence introduces a new type of
impurity in the plasma analogy of the so-called (111) ground state. That
enables us to easily derive meron charge distributions in the longdistance
limit and infer how the construction of a pair of merons
would look like in the same limit.

We also consider the same constructions in mixed, composite boson - composite
fermion, ground states \cite{srm}, proposed as a way to capture in the
ground state description the effect of quantum fluctuations at finite $d$, distance
between the layers. Charge screening of the single meron construction in
the plasma analogy of mixed states is almost without change with respect to the
(111) case. But the strength of the $ \ln(r)$ interaction in the plasma 
analogy between a pair of merons gets reduced, being propotional to
the density of bosons that decreases as a function of $d$. Because of, as detailed
below, a formal correspondence between the interaction laws between merons and
the interaction laws between impurities, this is very
suggestive of a mechanism, which (with composite fermion screening; see below)
might be responsible for a confinement weakening. Together with disorder 
the mechanism may lead
to the deconfinement believed to exist in the experiments \cite{eimac}.

In the following we will introduce plasma techniques \cite{meth} for the
 (111) state. Because of the unusual nature of the statistical model based
on the (111) state implied by the Laughlin prescription \cite{lau}, it is
not clear whether they are valid in this case, but we will show that indeed
they can capture the leading longdistance behavior. Let us begin with the most
obvious generalization of the Laughlin quasihole construction for the case
of the two-component, $\uparrow$ and $\downarrow$, (111) state,
\beq
\Psi(w) = \prod_{i = 1} (w - z_{\uparrow, i}) \Psi_{111}(z_{\uparrow},z_{\downarrow}),
\label{oc}
\eeq
where the (111) state is,
\beq
\Psi_{111}(z_{\uparrow},z_{\downarrow}) = \prod_{i < j} (z_{i, \uparrow} -
z_{j, \uparrow}) \prod_{k < l} (z_{k, \downarrow} -
z_{l, \downarrow}) \prod_{p , q} (z_{p, \uparrow} -
z_{q, \downarrow})
\eeq
(with omitted Gaussian factors). To get the charge distributions at point $r$ away
from the center $w$ of the excitation, we use an effective plasma expansion
 summing only contributions that can be symbolically
represented as the ones depicted in Fig. 1. 
\begin{figure}
\includegraphics[width= 0.90\columnwidth]{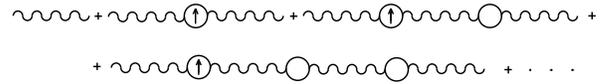}
\caption{Diagrammatic summation leading to the $\uparrow$ charge distribution away from
 an impurity}
\end{figure}
In the Figure,
 in the Fourier space, the wriggly line is the 2D Coulomb plasma interaction,
$ - \frac{2 \pi}{|\vec{q}|^{2}}$, and the vertices are: the empty circle of the
 value of the total density $n$ - the static structure factor of a Bose gas, and 
the circle with up arrow denotes the density $ n_{\uparrow}$ of the up particles.
 We probe the $\uparrow$ charge density at $r$ otherwise, if we probe $\downarrow$
 density the corresponding sum does not have the first contribution in Fig. 1. Therefore, to get
 the $\uparrow$ and $\downarrow$ charge distributions, we have to Fourier transform
$( \int d^{2} \vec{q} e^{i \vec{q} ( \vec{w} - \vec{r})} [ \cdots ])$ the following
expressions in which $ V(\vec{q}) = - \frac{2 \pi}{|\vec{q}|^{2}}$, 
\begin{equation}
\rho_{\uparrow}(q) = V(q) + \frac{V(q) n_{\uparrow} V(q)}{ 1 - n V(q)},
\label{gore}
\end{equation}
and
\begin{equation}
\rho_{\downarrow}(q) = \frac{V(q) n_{\uparrow} V(q)}{ 1 - n V(q)},
\label{dole}
\end{equation}
for $\uparrow$ and $\downarrow$ charge respectively. We immediately see
that the total charge is screened,
\begin{equation}
\rho_{c}(q) \sim  \rho_{\uparrow} + \rho_{\downarrow} = \frac{V(q)}{1 - n V(q)},
\end{equation}
and $ \lim_{q \rightarrow 0} \rho_{c}(q) = Const $, like in the usual
Laughlin quasihole case, but the pseudospin charge
$ \rho_{s}(q) \sim \rho_{\uparrow} - \rho_{\downarrow} = V(q)$ is unscreened
growing as $ \ln(r)$ (if $w = 0$) with distance $r$. Therefore the
capacitive energy
defined as
\begin{equation}
\label{ce}
E_{c} = \int d^{2} \vec{r} (\rho_{\uparrow} - \rho_{\downarrow})^{2},
\end{equation}
which is in the first approximation proportional to the energy to excite the
quasihole,
is proportional to (up to logarithmic factors) the area of the system.
This is the conclusion of the numerical study in \cite{ye}.
Therefore the plasma analogy is able to reproduce the main result of the
detailed investigation \cite{ye} (which helps us to eliminate from
further consideration the constructions of the form in Eq.(\ref{oc}) as
relevant excitations for the bilayer).
The agreement does not come as a surprise if we analyze more closely
the diagrams in Fig. 1. In them we are justifiably using the
screening properties of the charge channel,
which behaves as a plasma. Because of this, from now on, we will refer to
the statistical model
based on the (111) state as plasma.

In the second quantization formalism the meron excitation of
the pseudospin theory \cite{moo} that parallels the construction in Eq.(\ref{oc}) is
\begin{equation}
|\Psi_{m}(w=0)> = \prod_{m = 0}^{N - 1} ( c_{m + 1, \uparrow}^{\dagger}
+ c_{m, \downarrow}^{\dagger})|0>.
\label{nc}
\end{equation}
$c_{m,\sigma}^{\dagger}$'s create the lowest Landau level (LLL) states,
$\Phi_{m} = \frac{z^{m}}{\sqrt{2 \pi 2^{m} m!}} \exp\{- \frac{1}{4} |z|^{2}\},
m = 0, \ldots, N - 1$.
($N$ is the number of particles in the system.).
In the first quantization description of Eq.(\ref{nc}), $\uparrow$ orbitals
are shifted (in the expansion of the Slater determinant of a filled LLL)
in the following manner:
\begin{equation}
\Phi_{m}(z_{\uparrow}) \rightarrow \frac{z_{\uparrow}}{\sqrt{2(m + 1)}} \Phi_{m}(z_{\uparrow}).
\label{shift}
\end{equation}
This is a nontrivial change and can not be described by a simple multiplication operation on
the ground state like in Eq.(\ref{oc}). When $|z_{\uparrow}| \rightarrow \infty $,
more precisely when $ m = N - 1 \rightarrow \infty $ (i.e. $m$ is the last orbital
in the ground state) $ |\Phi_{m}(z_{\uparrow})|^{2}$ behaves, in the first approximation,
like a delta function, at $ |z_{\uparrow}| = \sqrt{2 (m + 1)}$, and, in this sense, we
can approximately take for multiplying $ z_{\uparrow}$ in Eq.(\ref{shift}),
$z_{\uparrow} = \exp\{i \phi\} \sqrt{2 (m + 1)}$. Then the excitation, very far from
the origin, looks like
\begin{equation}
\prod_{i = 1}^{N} 
\left[ \begin{array}{c} \exp\{i \phi_{i}\}  \\ 1 \end{array} \right]
\Psi_{111}(z_{\uparrow},z_{\downarrow})
\label{vortex}
\end{equation}
in the first quantization \cite{moo}. To get a change in the charge distribution away
from the origin we need further corrections to the limit in Eq.(\ref{vortex}).
We assume that they can be described by an expansion in the
 powers of $1 / (|z_{\uparrow}|)$, and we seek the coefficient of the
first correction by solving the following equation with $ |z| \equiv r$:
\begin{widetext}
\begin{equation}
\int^{\infty}_{\sqrt{2 (m + 1)}} dr \; r \; |z|^{2} |\Phi_{m}(z)|^{2} =
2 (m + 1) \int^{\infty}_{\sqrt{2 (m + 1)}} dr \; r \; |\Phi_{m}(z)|^{2} +
2 (m + 1)\; C \; \int^{\infty}_{\sqrt{2 (m + 1)}} dr \frac{ r |\Phi_{m}(z)|^{2} }{r}.
\label{cond}
\end{equation}
\end{widetext}
The condition and implied expansion are appropriate when we look for
charge density distributions of the state in Eq.(\ref{nc}).
In the limit $ m \rightarrow \infty$ we get $ C = 0.8 $ as can be seen in Fig. 2. 
In Eq.(\ref{cond}) the searched for correction is for the orbital
$\Phi_{m}$ right at the longdistance cut-off $R = \sqrt{2 (m + 1)} = \sqrt{2 N}$
(i.e., the radius of the system).
\begin{figure}
\includegraphics[width=\columnwidth]{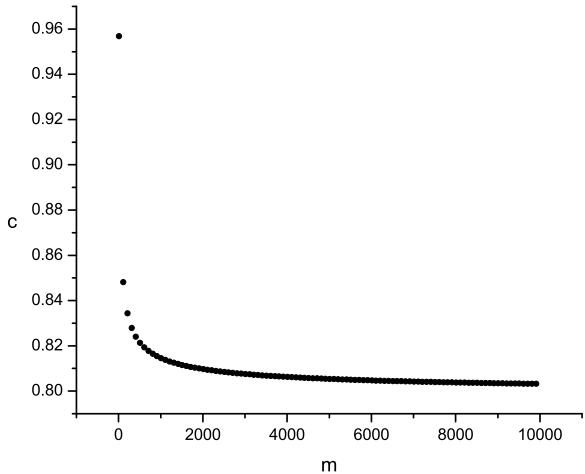}
\caption{Dependence of $C$ (Eq.(\ref{cond})) on $ m = N - 1$ ( the first value at $m = 10$ )}
\end{figure}
Therefore, in this approximation, in the approach with fixed up
and down number of particles, the $\uparrow$ charge density of the excitation
in Eq.(\ref{nc}) can be extracted from the following integral,
with $ |z_{\uparrow 1}| \equiv r $,
\begin{equation}
\rho_{w = 0}(r) \sim
\int d^{2}z_{\uparrow 2} \cdots \int d^{2}z_{\downarrow N}
\exp\{\sum_{i} \frac{C}{|z_{i \uparrow}|}\}
|\Psi_{111}(z_{\uparrow}, z_{\downarrow})|^{2},
\end{equation}
and analogously for the $\downarrow$ charge density.
In this way, in the (111) plasma, we consider a new type
of impurity which connects via the interaction
$ C / (|z_{i \uparrow}|) $ to the $\uparrow$ particles of the
plasma. {\em In this sense} we can propose the following longdistance
form of the meron excitation at some point $ w \neq 0 $ in general,
\begin{equation}
\prod_{i} \frac{z_{i \uparrow} - w}{|z_{i \uparrow} - w|}
\exp\{\sum \frac{C}{2 |z_{i \uparrow} - w|}\} \cdot
\Psi_{111}(z_{\uparrow},z_{\downarrow}).
\label{newcon}
\end{equation}
This construction can be easily generalized to the case when there are more
than one meron (of both vorticities). 

To get the charge distributions ($\uparrow$ and $\downarrow$) far away
from the center of the excitation, we use the same type of the approximation introduced
in the beginning (Fig. 1) with only one difference. Namely, we change the way
impurity connects to the plasma by switching from $ V(q) \sim \frac{1}{q^{2}}$ to
$V_{m}(q) \sim \frac{1}{q}$.
In this way $\rho_{\uparrow}(q) \approx \frac{1}{2} V_{m}(q)$ and
$\rho_{\downarrow}(q) \approx - \frac{1}{2} V_{m}(q)$ in the $q \rightarrow 0$ 
limit and for
$n_{\uparrow} = n_{\downarrow}$, resulting in $E_{c} \sim \ln R $, where $R$ is the radius of the system,
for the energy to excite a meron, in agreement with the XY model considerations
and pseudospin theory \cite{moo}.

 By considering the new impurities in the (111)
plasma and applying the plasma techniques, we can prove the usual XY model
logarithmic interactions between them, which is a result without an
obvious connection with the physics and XY model of the bilayer.
By considering also a pair of the
old impurities (that follow from the construction in Eq.(\ref{oc})), with same charge
and opposite vorticity, we can find that their interaction energy in the plasma
grows quadratically as a function of distance. It was found in \cite{ye}, in numerics,
that their real (capacitive) interaction energy behaves in the same way. This all
again shows that whenever we have an underlying bosonic analogy and corresponding quasiparticles,
like for the Laughlin states \cite{zh}, or transparent, like in the bilayer
case \cite{jye}, the corresponding plasmas have impurities with identical interaction
energy laws, up to the value of couplings, to the interactions among
 quasiparticles in the quantum Hall systems.

The mixed states proposed as the ground states \cite{srm} at finite (not small)
$d$  as mixtures of composite bosons of the (111)
state and composite fermions of the nearby phase of two decoupled Fermi-liquid-like
states can be expressed as
\begin{eqnarray}
\displaystyle
\label{talfja}
& \Psi_{o}=  {\cal PA} \{
\prod_{i<j}(z_{i\uparrow}-z_{j\uparrow})
\prod_{k<l}(z_{k\downarrow}-z_{l\downarrow})\prod_{p,q}(z_{p\uparrow}-z_{q\downarrow}) &  \nonumber\\
& \Phi_{f}^{\uparrow}(w_{\uparrow},\overline{w}_{\uparrow})\prod_{i<j}(w_{i\uparrow}-w_{j\uparrow})^2
\Phi_{f}^{\downarrow} (w_{\downarrow},\overline{w}_{\downarrow})\prod_{k<l}(w_{k\downarrow}-w_{l\downarrow})^2 & \nonumber\\ 
& \prod_{i,j}(z_{i\uparrow}-w_{j\uparrow})\prod_{k,l}(z_{k\uparrow}-w_{l\downarrow}) & \nonumber\\
& \prod_{p,q}(z_{i\downarrow}-w_{q\uparrow})\prod_{m,n}(z_{m\downarrow}-w_{n\downarrow})
\  \}. &  
\end{eqnarray}
$z$'s and $w$'s denote bosons and fermions respectively, 
$\Phi_{f}^{\sigma}, \sigma = \uparrow, \downarrow$ are two filled-Fermi-sea wave functions,
$ \cal{P}$ is the projection to LLL, and $\cal{A}$ is the antisymmetrizer for
bosons and fermions in each layer separately. 
The portion of composite fermions increases as $d$ increases.
Extracting the number of flux quanta -
the number of particles relations from Eq.(\ref{talfja}), we can find that the number
of up and down composite fermions must be the same. 
The mixed states are very close to
the exact diagonalization ground states.
In the following we will apply on them the weakly-screening plasma
approach \cite{meth}, which, again, in the longdistance approximation, was able to
reproduce the basic physics of the Fermi-liquid-like composite fermion states. Because
of the presence of $ \Phi_{f}$'s, any vertex representing connection through composite
fermions is effectively the static structure factor of free Fermi gas i.e.
$ s_{\sigma}(q) \sim q$ in the small momentum limit.

We can consider excitations of the type described by Eq.(\ref{newcon}) in a mixed state.
Let us consider the following dipole construction of the introduced excitation:
\begin{widetext}
\begin{equation}
\prod_{i} \frac{z_{i \uparrow}}{|z_{i \uparrow}|}
\prod_{j} \frac{w_{j \uparrow}}{|w_{j \uparrow}|}
\prod_{k} \frac{\bar{z}_{k \downarrow}}{|z_{k \downarrow}|}
\prod_{l} \frac{\bar{w}_{l \downarrow}}{|w_{l \downarrow}|}
\exp\{\frac{C}{2} (\sum_{i} \frac{1}{|z_{i \uparrow}|} + \sum_{j} \frac{1}{|w_{j \uparrow}|} - 
 \sum_{k} \frac{1}{|z_{k \downarrow}|} - \sum_{l} \frac{1}{|w_{l \downarrow}|}) \} \; \Psi_{o}.
\label{dipole}
\end{equation}
\end{widetext}
We will take that the number of $\uparrow$ and $\downarrow$ bosons is the same and 
neglect the antisymmetrizer in $\Psi_{o}$ . Then, the charge
distribution of $\uparrow$ charge can be found, first considering bosonic
part, which has only a single connection and contribution $\sim V_{m}(q)$ because of
the alternating sign in Eq.(\ref{dipole}) and fermionic part that we
symbolically depicted in Fig. 3. 
\begin{figure}
\includegraphics[width= 0.90\columnwidth]{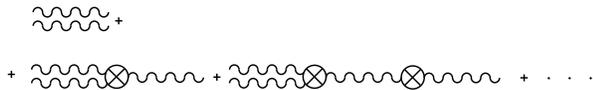}
\caption{Diagrammatic summation leading to the fermionic part of the
charge distribution away from the dipole (Eq.(\ref{dipole}))}
\end{figure}
In Fig. 3 the doubly wriggly line denotes $V_{m}(q)$,
and the crossed circle, twice the static structure factor, $s_{\uparrow}(q)$, of the
Fermi gas.
 Though simple, the final contributions are results of massive cancellations that
follow from different interaction signs in the way quasiparticles connect to the plasma.
Examining the contributions, we find that despite the presence of the
composite fermions and their screening, the $\uparrow (\downarrow)$ charge
distributions (effectively bosonic) are again propotional to $ + (-) \frac{1}{r}$,
leading to the infinite energy requirement for the excitation in the
thermodynamic limit.
The presence of the composite fermions signifies the quantum fluctuations as
the distance between the layers increases. Nevertheless, 
we can again conclude, 
 the excitations in the
pseudospin channel (no charge, only vorticity), like the one we considered,
retain their meron confinement property \cite{moo}.

This conclusion is corraborated by a calculation of the plasma interaction
among the meron pair of opposite vorticity but the same charge in a mixed state.
We state the final result,
\begin{widetext}
\begin{equation}
V_{int}(q) =
\frac{V_{m}^{2}(q) V(q) (n_{\uparrow} n_{\downarrow} + n_{\uparrow} s_{\uparrow}(q) + 
n_{\downarrow} s_{\downarrow}(q))}{1 - V(q) (n + 2 s(q))} +
\frac{V_{m}^{2}(q) s^{2}(q)}{1 - V^{2}(q) (2 s(q))^{2}}
\; \{ n V^{2}(q) +
\frac{n^{2} + n (2 s_{\uparrow}(q) + 2 s_{\downarrow}(q))}{1 - V^{2}(q) (2 s(q))^{2}}
\; V^{3}(q) \},
\end{equation}
\end{widetext}
where $s_{\uparrow}(q)=s_{\downarrow}(q)=s(q)$ and $n_{\uparrow}, n_{\downarrow},$ and
$n = n_{\uparrow} + n_{\downarrow}$ denote {\em bosonic} densities. This can be
obtained straightforwardly with the help of diagrams. In the $q \rightarrow 0$ limit
we have,
\begin{equation}
V_{int}(q) \rightarrow (-) V_{m}^{2} \frac{n_{\uparrow} n_{\downarrow}}{n} +
(2 \frac{n_{\uparrow} n_{\downarrow}}{n^{2}} - \frac{1}{2}) 
V_{m}^{2}(q) s(q),
\end{equation}
i.e. the leading is the attractive $\ln(r)$ interaction, and the correction is a
 $\frac{1}{r}$ interaction due to the screening by composite fermions that
vanishes in the $n_{\uparrow} = n_{\downarrow}$ case \cite{end}. This is a 
result in the formal setting of plasma analogy, but very likely, due to the mentioned
correspondence, also
a relevant conclusion for the interaction between  two merons in the quantum Hall
system. 
Then please note again that $n_{\uparrow}, n_{\downarrow},$ and $n$, are
not overall densities but reduced, due to the presence of fermions, bosonic densities.
Therefore, though the type of interaction $(\ln(r))$ stays the same, the coupling strength
is weaker due to its proportionality to the density of bosons.

Certainly, it is appropriate to check  the amounts of the screening charges of
 a single meron construction in a mixed state that is 
 the generalization
of the construction in Eq.(\ref{newcon}).
Again with the help of diagrams, they can be easily found, and we will
just state their limiting, $q \rightarrow 0$, behavior:
\begin{equation}
\rho_{\uparrow}(q) \rightarrow \frac{V_{m}(q)}{2} + \frac{n_{\uparrow} - n_{\downarrow}}{n}
 V_{m}(q),
\end{equation}
and,
\begin{equation}
\rho_{\downarrow}(q) \rightarrow \frac{V_{m}(q)}{2} + \frac{ - 2 n_{\downarrow}}{n}
 V_{m}(q).
\end{equation}
Therefore, in the $n_{\uparrow} = n_{\downarrow}$ case, the limits do not
differ  from the case without composite fermions.

M.V.M. thanks D.A. Huse for an inspiring conversation, N. Read, E.H. Rezayi, and S.H. Simon
 for previous collaborations, and Lucent Bell Labs for their hospitality during the time
when this work was initiated.
The work was supported by
Grant No. 1899 of the Serbian Ministry of Science.


\begin{references}
\bibitem{end}Very likely this correction is not a final result because of the known
difficulties of the weakly-screening plasma approach \cite{meth} in
generating nonanalytic ( $\sim |q|^{n}, n$ - odd) correction to a leading behavior.
\end{references}
\end{document}